\documentclass[prl,10pt,aps,twocolumn,amsmath,amssymb,superscriptaddress]{revtex4-1}
\pdfoutput=1
\usepackage{hyperref}
\usepackage{graphicx}
\usepackage{color}
\usepackage{multirow}
\usepackage{amsmath}
\usepackage{array}
\usepackage{booktabs}
\usepackage{float}
\usepackage{multirow}
\usepackage{cleveref}

\newcommand{\be}{\begin{equation}}
\newcommand{\ee}{\end{equation}}
\newcommand{\bea}{\begin{eqnarray}}
\newcommand{\eea}{\end{eqnarray}}
\def\c#1{~\cite{#1}}
\def\f#1{Fig.~\ref{#1}}

\def\beq{\begin{equation}}
\def\eeq{\end{equation}}

\begin{document}

\title{Bypassing sluggishness: SWAP algorithm and glassiness in high dimensions}

\author{Ludovic Berthier}
\affiliation{Laboratoire Charles Coulomb (L2C), University of Montpellier, CNRS, Montpellier, France}
\author{Patrick Charbonneau}
\affiliation{Department of Chemistry, Duke University, Durham, North Carolina 27708, USA}
\affiliation{Department of Physics, Duke University, Durham, North Carolina 27708, USA}
\author{Joyjit Kundu}
\affiliation{Department of Chemistry, Duke University, Durham, North Carolina 27708, USA}
\email{joyjitkundu032@gmail.com}

\date{\today}

\begin{abstract}
The recent implementation of a swap Monte Carlo algorithm (SWAP) for polydisperse mixtures fully bypasses computational sluggishness and closes the gap between experimental and simulation timescales in physical dimensions $d=2$ and $3$. Here, we consider suitably optimized systems in $d=2, 3,\dots, 8$, to obtain insights into the performance and underlying physics of SWAP. We show that the speedup obtained decays rapidly with increasing the dimension. SWAP nonetheless delays systematically the onset of the activated dynamics by an amount that remains finite in the limit $d \to \infty$. This shows that the glassy dynamics in high dimensions $d>3$ is now computationally accessible using SWAP, thus opening the door for the systematic consideration of finite-dimensional deviations from the mean-field description. 
\end{abstract}

\maketitle

{\em Introduction --} A glass emerges when a supercooled liquid passed its crystallization point becomes so sluggish that it falls out of equilibrium. Upon cooling or increasing packing fraction, the dynamics of glass formers exhibits a marked slowdown beyond the dynamical onset, thus making this outcome inescapable\c{general-glass,general-glass2}. In mean-field descriptions, the structural relaxation time exhibits a power-law divergence at the dynamical transition\c{ludo_rmp}. In any finite dimension, although activated processes wash out this transition, the rapid growth of the associated relaxation time nonetheless impedes equilibration of low-temperature or high-density liquids. Standard simulation protocols, in particular, do not easily explore the regime beyond the dynamic transition, because structural relaxation is already too sluggish. 

The application of the swap Monte Carlo algorithm (SWAP), which exchanges the identity of pairs of particles, to complex mixtures sidesteps this difficulty\c{swap-prl,swap-karmakar,swap-parisi}. By considering systems with, for instance, a continuous size polydispersity one can follow the equilibrium liquid up to unprecedented high packing fractions or low temperatures. Tuning the range and functional form of polydispersity provides systems for which the sampling efficiency of swap moves is maximal within the liquid state, while remaining robust against crystallization and fractionation\c{ludo_prx}. For properly chosen polydispersities in $d= 3$ this procedure has recently provided a speedup of at least $10^{10}$ compared to standard dynamics, matching the experimental timescales\c{ludo_prx,sho_pnas}, and in $d=2$ it has given access to timescales that are truly cosmological\c{sho2018}. 
This computational progress has triggered the exploration of new glass physics in computer simulations, notably low-temperature anomalies\c{camille2017,lijin2018}, the Gardner transition\c{ludo_pnas2016,camille2017}, the rheology of glasses\c{misaki2018}, the extension of the jamming line\c{misaki_scipost2017}, and the ultrastability of vapor-deposited glasses\c{flenner2017}.

The efficiency of SWAP has also triggered theoretical activity aimed at better understanding its physical origin and its physical implications for the glass transition\c{wyart_cates}.
Ikeda \textit{et al.}\c{ikeda2017} present a replica calculation of a mean-field glass model proposing that SWAP and physical dynamics are ruled by distinct dynamical transitions. A qualitatively similar result is obtained by 
Szamel who obtains two dynamical transitions for the two dynamics\c{szamel2018}. Brito \textit{et al.}\c{brito2017} obtain a similar result, and interpret the dynamical transition as an onset of mechanical rigidity that is again shifted by SWAP. Finally, Berthier \textit{et al.}\c{ludoswap} argue that the onset of thermal activation past the dynamical transition is also considerably affected by SWAP. There is thus a general consensus that SWAP can delay the dynamical transition by an amount that is system dependent, and can speedup the dynamics even past the avoided dynamical transition. 

However, because dynamical transitions are avoided in any finite 
$d$\c{Charbonneau:2017}, other physical processes might also explain the dramatic change in dynamics. In particular, structural imperfections closely tied to local geometry\c{paddy_review}, which are putatively important in the dynamics of low-dimensional glass formers, could impact SWAP efficiency. Distinguishing one contribution from the other can be achieved by considering how SWAP performance evolves with increasing $d$. A non-vanishing SWAP efficiency in the limit of $d\to \infty$ or a perturbative correction in $1/d$ would suggest that the mean-field dynamical transition is indeed shifted, while an exponential suppression would suggest that nonperturbative features associated with geometry dominate. Because numerical work on SWAP has thus far only been concerned with physical dimensions, $d=2$ and 3, distinguishing between these scenarios is not currently possible. 

Resolving this question would not only shed light on the physical origin of the glassy slowdown, but help devise novel algorithms that further bypass it. Interestingly, side-stepping the mean-field dynamical threshold could also be key to general algorithmic improvements in hard problems, such as statistical inference, high-dimensional optimization and deep learning\c{lenka2016}. A fundamental grasp of the effectiveness of SWAP dynamics could thus bolster advances far beyond the problem at hand. More immediately, if one could generically push the current limitations of high $d$ simulations, crucial questions in glass physics could be tackled\c{Charbonneau:2017,reichman2009,sastry_jcp}. 
In this work, we study the dynamics of suitably optimized polydisperse mixtures of hard spheres in various spatial dimensions, so as to systematically approach the mean-field, $d\rightarrow\infty$ description, and provide microscopic insight into the underlying physics and computational efficiency across a broad range of dimensions. 

\begin{figure}
\includegraphics[width=0.94\linewidth]{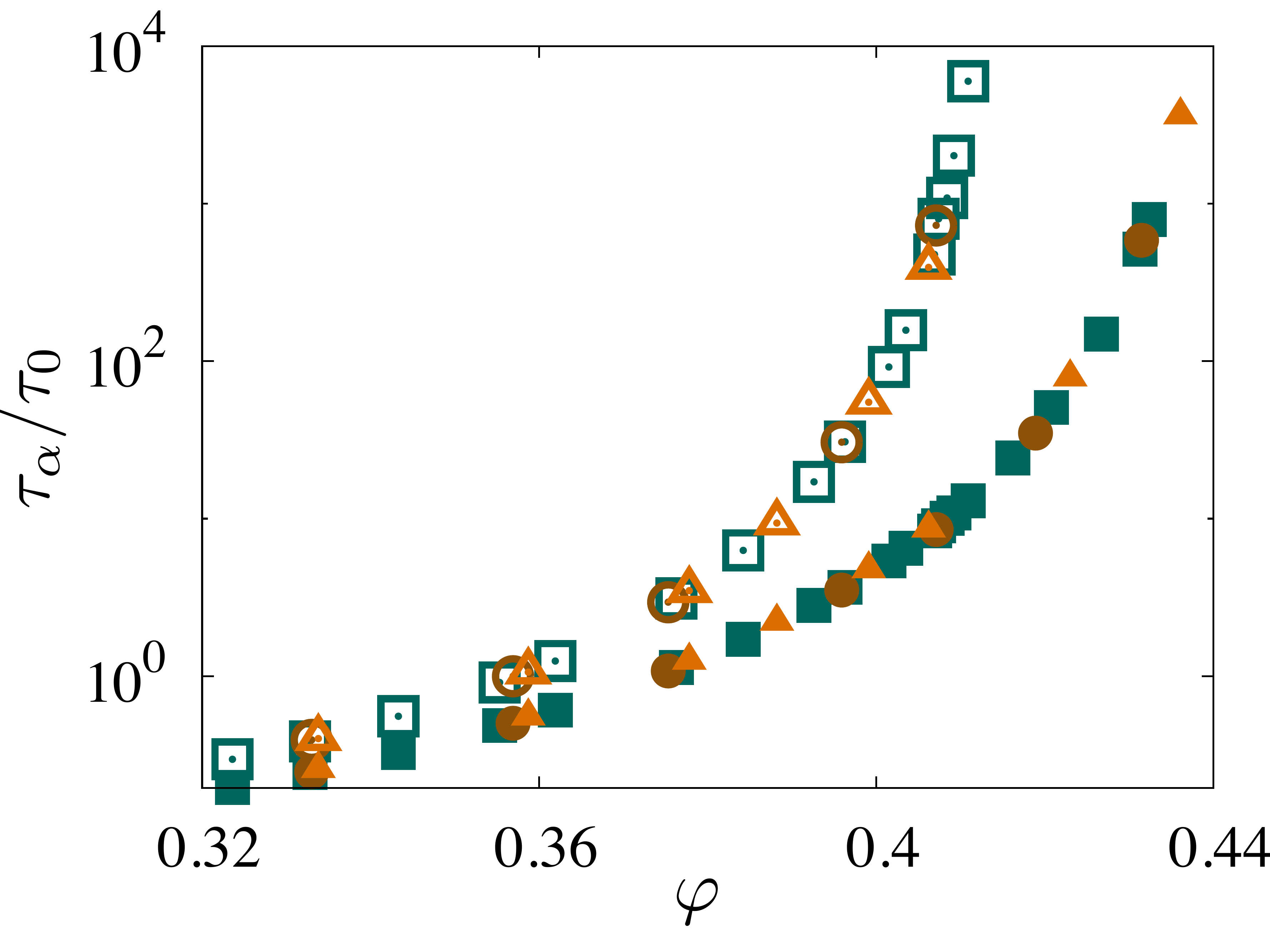}
\caption{Structural relaxation time of standard (open symbols)  and SWAP dynamics (solid symbols)  for various particle size distributions $P(\sigma)$ with $\Delta=10\%$ in $d=4$: flat (circles), $1/\sigma^3$ (squares), and $1/\sigma^4$ (trigangles). For a given $\Delta$ both dynamics are unaffected by the functional form of $P(\sigma)$.}
\label{fig01}
\end{figure}

{\em Simulation Model--} We consider size polydisperse systems with $N$ hard spheres in a hypercubic box of constant volume $V$, under periodic boundary conditions in $d=2, 3 \cdots, 8$. The size distribution function has the form, $P(\sigma)=K/\sigma^3$, with normalization constant  $K$ for $\sigma \in [\sigma_{\rm min}, \sigma_{\rm max}]$, where $\sigma_{\rm min}$ and $\sigma_{\rm max}$ are the minimum and the maximum diameter values, respectively. The average diameter $\bar{\sigma}=\int_{\sigma_{\rm min}}^{\sigma_{\rm max}} P(\sigma) \sigma$ sets the unit of length, and the standard deviation of the size distribution, $\Delta$, quantifies the degree of polydispersity (see Simulation details and model parameters in\c{note}). For a fixed $\Delta$, this specific choice of size distribution function does not significantly affect the system dynamics. Figure~\ref{fig01}, which explicitly compares the dynamics at fixed $\Delta$ and various $P(\sigma)$ in $d=4$, confirms that $\Delta$ is the most relevant variable. Our analysis should therefore be reasonably independent of the specifics of the model studied.

Standard and SWAP simulations are run for different $\Delta$ and $d$. Both dynamical protocols include basic single-particle translational moves along a vector randomly drawn within a $d$-dimensional hypercube of side $\delta \ell$; SWAP includes additional diameter exchanges between two randomly chosen particles, attempted with probability $p=0.2$ (setting $p=0$ recovers standard dynamics). While $0<p\lesssim0.2$ monotonically increases sampling efficiency, for $p\gtrsim 0.2$ efficiency saturates, and hence additional swap moves wastefully slow down simulations\c{ludo_prx}. For each volume fraction $\varphi$, the pressure $P$ is measured using pair correlations\c{santos2002,yuste2005}, to compute the unitless reduced pressure, $Z = \beta P/\rho$, for the number density $\rho=\varphi/\bar{V}_\mathrm{d}$ with  $\bar{V}_\mathrm{d}$ being the average volume of a $d$-dimensional hypersphere. 

Equilibration is assessed by the complete decay of the self-part of the particle-scale overlap function
\begin{equation}
Q(t)=\frac{1}{N}\displaystyle \sum_{i=1}^{N} \Theta(a-|\mathbf{r}_i(t)-\mathbf{r}_i(0)|),
\end{equation} 
where $\Theta$ is a step function and $a=0.3\bar{\sigma}$ is a microscopic length chosen to be close to the typical particle cage size. The associated structural relaxation time, $\tau_\alpha$, is defined such that $Q(\tau_{\alpha})=e^{-1}$. We define the relaxation time for both the standard ($\tau_\alpha^{\rm std}$) and SWAP ($\tau_\alpha^{\rm swap}$) dynamics. In all dimensions studied, SWAP equilibrates systems far beyond what is computationally accessible with standard Monte Carlo, and we thus first equilibrate systems using SWAP before measuring properties of the dynamics without it.

\begin{figure}
\includegraphics[width=\columnwidth]{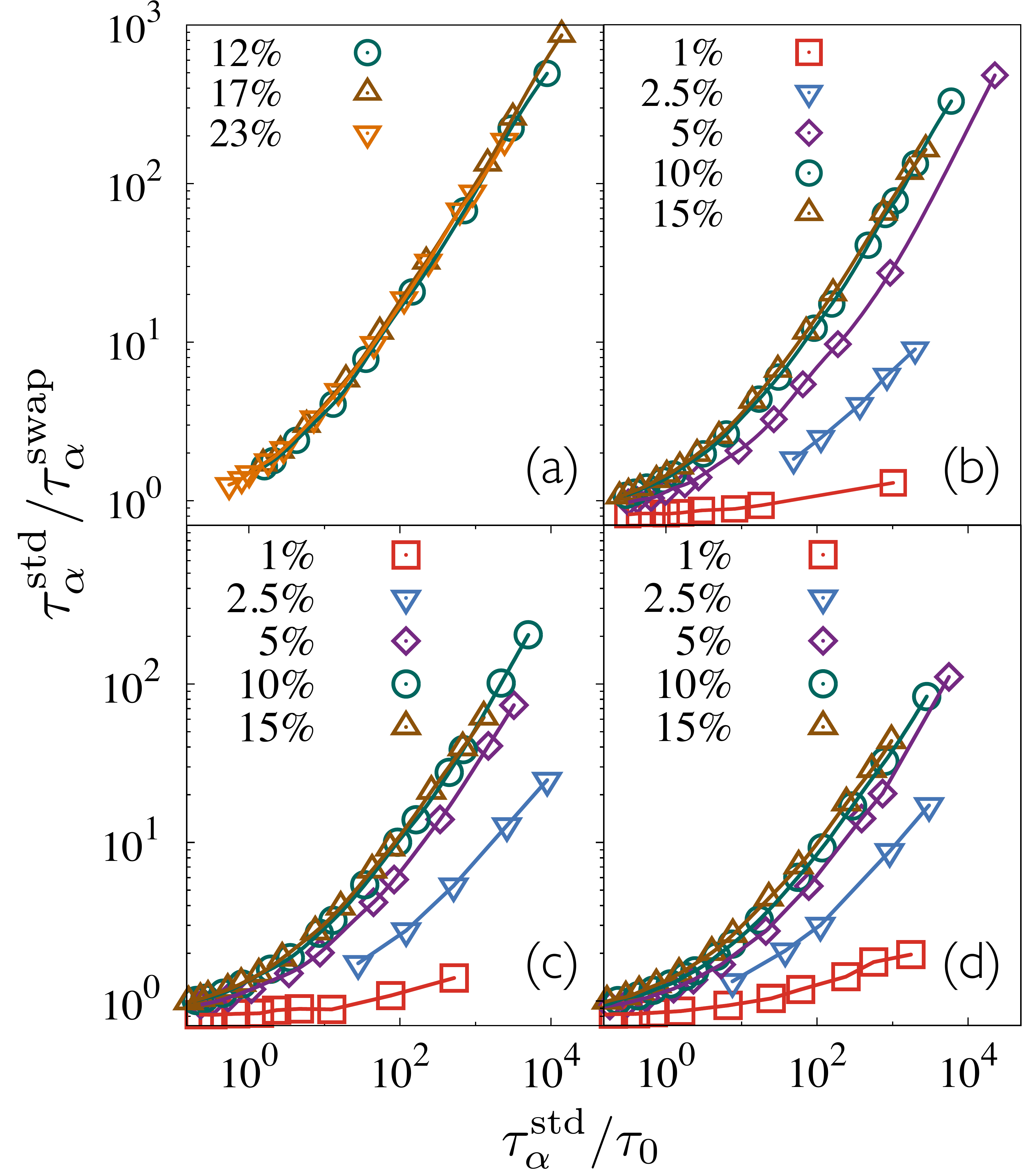}
\caption{SWAP efficiency, $\tau^{\rm std}_\alpha / \tau^{\rm swap}_\alpha$, as a function of the relaxation time of the standard dynamics (representing the sluggishness) for different polydispersities $\Delta$ in (a) $d=3$, (b) $d=4$, (c) $d=5$, and (d) $d=6$. Sluggish dynamics at low $\Delta$ cannot be reached in $d=3$ because crystallization interferes. In all $d$, SWAP performs better as $\Delta$ increases, and saturates at larger $\Delta$.}
\label{fig02}
\end{figure}

\begin{figure*}
\includegraphics[width=\linewidth]{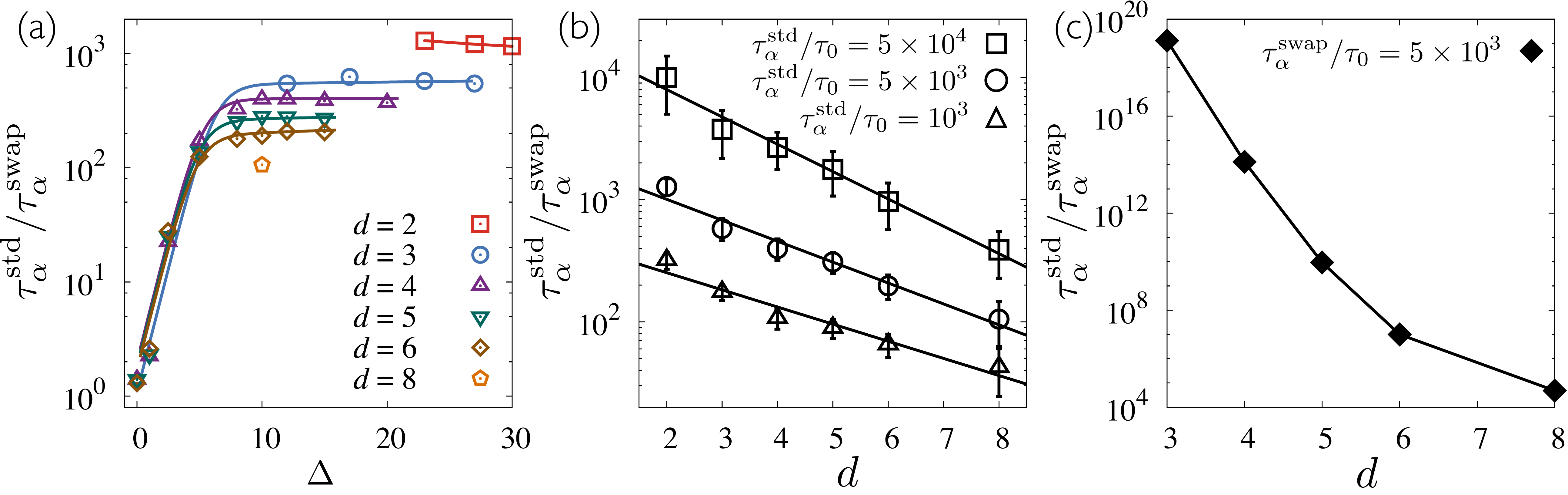}
\caption{(a) Evolution of SWAP efficiency, $\tau^{\rm std}_\alpha / \tau^{\rm swap}_\alpha$, with $\Delta$, measured at $\tau^{\rm std}/\tau_0=5\times 10^3$. A saturation to a plateau value occurs at lower $\Delta$ and at a lower plateau value as $d$ increases. (b) Same quantity measured at $\tau_\alpha^{\rm std}/\tau_0=5\times 10^4$, $5\times 10^3$ and $10^3$ as a function of $d$. Exponential fits are denoted as solid lines. (c) Estimated SWAP efficiency at $\tau_\alpha^{\rm swap}/\tau_0=5\times 10^3$, which is roughly the edge of the computationally accessible regime using SWAP. The dramatic speedup in $d=3$ decreases rapidly with $d$, but remains larger than 4 orders of magnitude in $d=8$.} 
\label{fig03}
\end{figure*}

{\em Results--} In physical dimensions, crystallization  competes with equilibration of deeply supercooled liquids\c{valerini2011}. For instance, for $\Delta \lesssim 8\%$ in $d=3$ crystallization at high $\varphi$ is unavoidable. For $d>3$, by contrast, crystallization does not interfere with the metastable fluid phase even for arbitrarily low $\Delta$. The nucleation time  at finite $\Delta$ in $d>3$ is thus as equally out of computational reach as it is for monodisperse systems ($\Delta=0$)~\cite{skoge2006,fortini2009,meel_pre2010}. In all $d$, however, size fractionation may take place at high $\Delta$ and $\varphi$. In $d=3$, fractionation appears at $\Delta \gtrsim 10\%$, which helps crystallization~\cite{truskett,daniele}. In practice, this only happens when SWAP is used~\cite{ludoswap}, because composition fluctuations leading to fractionation are then much faster. SWAP thus not only accelerates the sampling of the metastable fluid, but also changes the glass-forming ability of the system and forces the use of $\Delta > 20\%$ in $d=3$. In $d=4$, by contrast, fractionation only appears at $\Delta \gtrsim 15\%$ for $\varphi \gtrsim 0.43$, and is further suppressed at higher $\Delta$ (see Dynamic and static observables in\c{note}). For each $d$, a $\Delta$ window, within which SWAP efficiency is reasonably good and fractionation (with or without crystallization) does not interfere, can thus be found. Qualitative and even quantitative aspects of the standard Monte Carlo dynamics are otherwise not remarkably affected by changing $\Delta$, as expected from previous studies of naturally polydisperse systems, such as colloidal suspensions\c{weeks2012}. 

A strong dependence of the SWAP dynamics on $\Delta$ is observed in the dynamically sluggish regime, beyond the onset of slow diffusion at $\varphi_0$ (\f{fig02}(a)-(d)). As an illustration, we consider the evolution of the SWAP efficiency ratio, $\tau_\alpha^{\rm std}/\tau_\alpha^{\rm sw}$ measured at a fixed $\tau_\alpha^{\rm std}/\tau_0$, with $\tau_0\equiv\tau_\alpha(\varphi_0)$. In \f{fig03}(a), we specifically consider $\tau_\alpha^{\rm std}/\tau_0=5 \times 10^3$, but the results are qualitatively robust for $\tau_\alpha^{\rm std}/\tau_0>1$ (see Dynamic and static observables in\c{note}). At low $\Delta$, SWAP dynamics is indistinguishable from standard dynamics and its efficiency increases monotonically. This efficiency, however, essentially saturates beyond a certain $\Delta$, resulting in its overall sigmoidal growth.  We empirically fit the results to a generalized logistic function, $S(\Delta)=A\exp(a\Delta)/(B+\exp(b\Delta))$, with fit parameters $A$, $a$, $b$, and $B$, to quantify the crossover polydispersity, $\Delta_0$, defined such that $S(\Delta_0)=0.9A$. We obtain $\Delta_0 \approx 10\%$ in $d=3$, $\approx 7.5\%$ in $d=4$, and $\approx 7\%$ in $d=5$ and $\approx 6.5\%$ in $d=6$. In $d=2$ and $3$, the trend is almost hidden by crystallization, and had gone unnoticed in previous work. The shrinking of $\Delta_0$ with increasing $d$ is nonetheless very clear. No theoretical framework formally predicts the saturation with $\Delta$ and the associated scaling with dimension. Physically, we interpret these results as follows. The amplitude of particle size fluctuations, which help uncage particles in SWAP dynamics, increase with $\Delta$, which accounts for the initial growth of efficiency with $\Delta$. The diffusion of particle diameters beyond a typical size, however, itself becomes slower than the  structural relaxation when $\Delta$ is large, because diameter and position dynamics are intimately coupled\c{ludo_prx}. Increasing $\Delta$ thus no longer improves SWAP efficiency, and this saturation develops earlier in larger $d$, where the vibrational dynamics (or, loosely speaking, caging) itself occurs over a length-scale decreasing with $d$. 

The most remarkable feature of the efficiency results is the weakening of SWAP performance with increasing $d$.  \f{fig03} (b) shows that the efficiency decays rapidly with increasing $d$ (nearly exponentially, at least up to $d=8$) for various $\tau_\alpha^{\rm{std}}/\tau_0$. The decay of SWAP performance becomes more prominent when estimated beyond the accessible regimes of the standard dynamics, such as where $\tau_\alpha^{\rm swap}/\tau_0=5\times 10^3$ -- see \f{fig03} (c) (and Dynamic and static observables in\c{note}). 

In order to examine explicitly whether this strong suppression is due to non-perturbative effects of not, we consider how SWAP impacts the avoided mean-field dynamical transition, $\varphi_\mathrm{d}$. We estimate $\varphi_\mathrm{d}$ for both standard dynamics and SWAP by fitting the growth of the relaxation time to the critical scaling form, $\tau_\alpha \propto (\varphi_\mathrm{d}-\varphi)^{-\gamma}$~\cite{Charbonneau:2017} (see Mode coupling analysis in\c{note}). As expected\c{patrick_pnas2014}, this scaling form captures the data increasingly well as $d$ increases. In $d=2$, it does not have a good regime of validity, but its validity eventually reaches up to three decades in the computationally accessible regime. We find that $\gamma$ is fairly insensitive to both dimension~\cite{Charbonneau:2017} and polydispersity\c{fuch2010}. Three features of the results are particularly noteworthy. First, collapsing $\tau_\alpha^{\rm std}/\tau_0$ by rescaling $\varphi/\varphi_{\rm d}^{\rm std}$ clearly reveals that SWAP postpones the putative dynamical transition in all dimensions--\f{fig04}(a).  
Second, while $\varphi_\mathrm{d}$ monotonically grows with polydispersity\c{biazzo2009}, its relative impact,  $(\varphi_\mathrm{d}^{\rm swap}-\varphi_\mathrm{d}^{\rm std})/\varphi_\mathrm{d}^{\rm{std}}$ eventually plateaus on a scale  consistent with the estimates for $\Delta_0$--see \f{fig04}(b). This suggests that the shift of dynamical transition is directly correlated with the SWAP efficiency, as both quantities evolve similarly with $\Delta$. Third, the plateau height, $h\equiv(\varphi_\mathrm{d}^{\rm swap}-\varphi_\mathrm{d}^{\rm std})/\varphi_\mathrm{d}^{\rm{std}}$ at the maximum polydispersity considered in \f{fig04}(b), decays to a nonzero value $(\approx 0.037)$ with correction that scales with dimension as $\sim 1/d$. Our results thus suggest that the gain in SWAP efficiency survives in the limit $d\to \infty$, and that perturbative corrections survive all the way down to $d=3$, independently of non-perturbative effects. 

\begin{figure}
\includegraphics[width=\columnwidth]{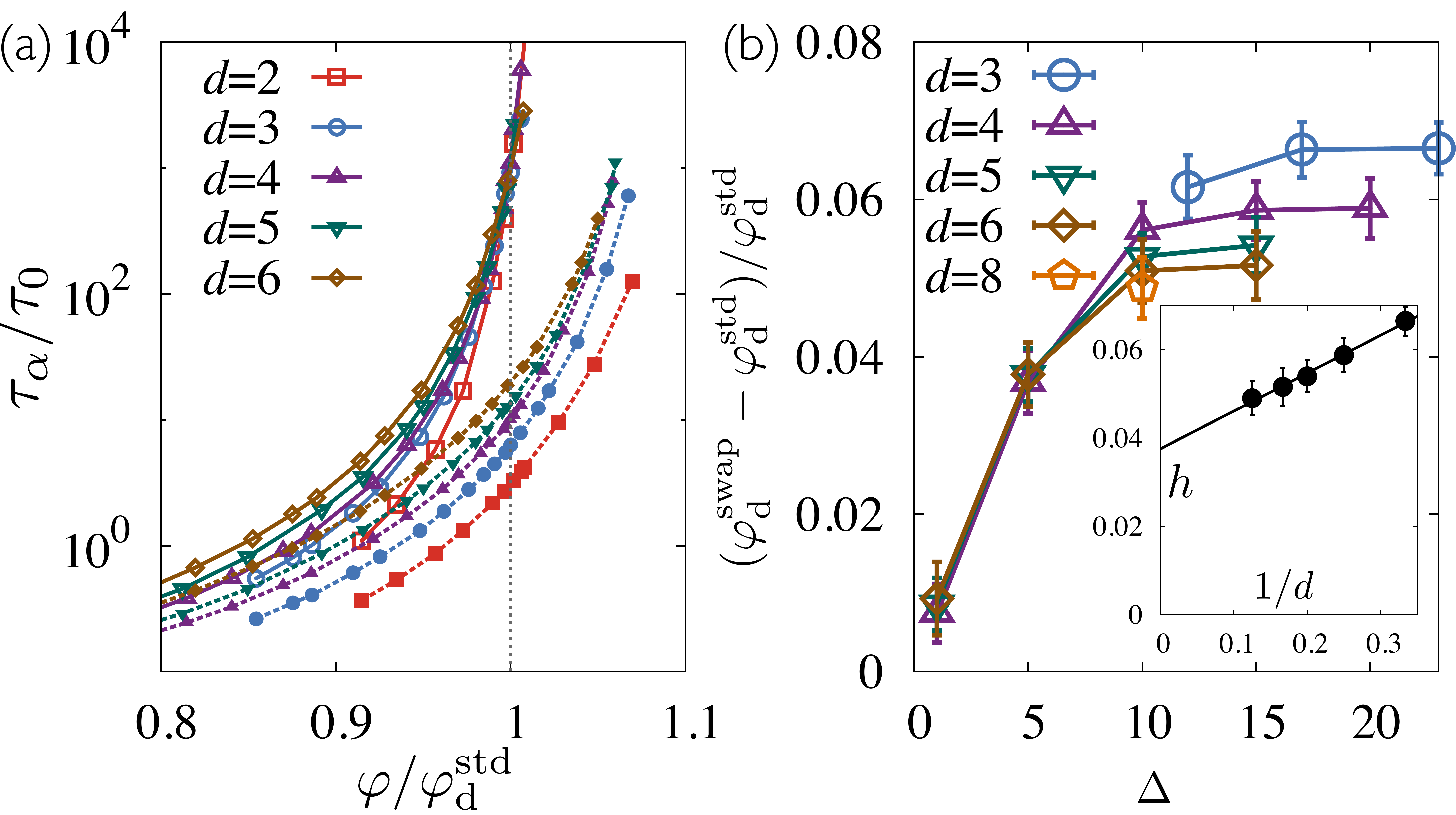}
\caption{(a) The structural relaxation time $\tau_\alpha^{\rm std}$ (open symbols) collapses for different $d$ upon rescaling $\varphi/\varphi_{\rm d}^{\rm std}$. The gap between SWAP and standard dynamics shrinks and saturates with increasing $d$. (b) The relative gap between the dynamical transition density between the two dynamics $(\varphi_\mathrm{d}^{\rm sw}-\varphi_\mathrm{d}^{\rm std})/\varphi_\mathrm{d}^{\rm std}$ also saturates at high $\Delta$, (inset) and the saturation height asymptotically approaches a nonzero value ($\approx 0.037$) as $\sim 1/d$.} 
\label{fig04}
\end{figure}

How can one explain the relatively rapid suppression of swap efficiency despite of the slow decay of the density gap $(\varphi_\mathrm{d}^{\rm swap}-\varphi_\mathrm{d}^{\rm std})/\varphi_\mathrm{d}^{\rm{std}}$ to a nonzero value? While the relative increase of $\varphi_\mathrm{d}$ is qualitatively consistent with mean-field treatments in $d=3$\c{ikeda2017,szamel2018}, the saturation and the asymptotic behavior of the gap with $d$ were not anticipated. 
Plugging this result into the critical scaling forms $\tau_\alpha = A(\varphi_\mathrm{d}-\varphi)^{-\gamma}$, we obtain an approximate expression for the efficiency ratio 
\be
\tau_\alpha^{\rm std} / \tau_\alpha^{\rm swap} \approx 
\tau_\alpha^{\rm std} ( h \varphi_d^{\rm std} )^\gamma
\ee
For a given value of $\tau_\alpha^{\rm std}$, the key contribution to the efficiency gain therefore arises from the term 
$( h \varphi_\mathrm{d}^{\rm std} )^{\gamma}$. Because asymptotically $\varphi_\mathrm{d}^{\rm std}\sim d~2^{-d}$\c{Charbonneau:2017}, this gain decreases rapidly with increasing $d$ -- qualitatively consistent with Figs.~\ref{fig03} (b, c) and Fig.~\ref{fig04}(b). Because $\tau_\alpha$ diverges upon approaching $\varphi_\mathrm{d}$ in high dimension, however, one should always be able to identify sufficiently sluggish systems for SWAP to speed up sampling. In intermediate dimensions, the approach remains sufficiently productive to obtain equilibrium configurations much beyond the dynamical transition of the standard dynamics. Figure~\ref{fig03} (c) provides a rough estimates of how useful SWAP can be in accessing regimes that are not accessible by the standard dynamics in high dimensions. For instance, in $d=8$ a speed up of roughly $10^4$ should remain computationally achievable. 

\paragraph{Conclusion --}
We have shown that SWAP improves sampling in dimensions $d \geq 2$ by generically delaying the dynamical transition that indicates the emergence of activated dynamics in the standard dynamics. This finding in itself does not directly reveal the microscopic nature (dynamic or thermodynamic) of the standard dynamics in the regime $\varphi_d^{\rm std} < \varphi < \varphi_d^{\rm swap}$, where SWAP provides most of its dynamic speedup, but offers a platform for assessing this question in the future.
Because the gap between the dynamical transition of the standard and the SWAP dynamics remains finite in the limit  $d\rightarrow\infty$, SWAP can efficiently be used to study pure glass physics in reasonably large dimensions, far from the regime in which significant local structure\c{paddy_review} or orientational ordering\c{tanaka2018} might interfere. In other words, although caging imperfections go away exponentially quickly with increasing dimension, SWAP can still break cages in high $d$.  Even within this analysis, the two-dimensional speedup is remarkably large, and 
techniques specifically tailored to identify local structural weaknesses, (e.g., \c{paddy_prx,paddy_nat_comm, liu_nat_phys, liu_pnas}) might thus help obtain additional microscopic insights. More generally, our observations suggest that the standard dynamical transition might not be as strong an algorithmic constraint as previously conceived in problems ranging from physics to information theory. If a proper sampling scheme can be devised and exploited in those problems, other stunning algorithmic advances might thus be within reach.

\begin{acknowledgments}
We thank S. Yaida, M. Ozawa and F. Zamponi for useful discussions.  J. K., L. B. and P. C. acknowledge support from the Simons Foundation grant (\#454933, Ludovic Berthier, \# 454937, Patrick Charbonneau). Most simulations were performed at Duke Compute Cluster (DCC). J.K. thanks Tom Milledge for helping with the usage of DCC. P.C. and J.K. also thanks Extreme Science and Engineering Discovery Environment (XSEDE), supported by National Science Foundation grant number ACI-1548562, for computer time.
\end{acknowledgments}

\onecolumngrid
\clearpage

~\\
\begin{center}
{\noindent\Large{Supplemental Information}}\\
\noindent \Large for\\
{\noindent \Large ``Bypassing sluggishness: SWAP algorithm and glassiness in high dimensions''}\\
~\newline
{\noindent \normalsize Ludovic Berthie$\mbox{r}^{1}$, Patrick Charbonnea$\mbox{u}^{2,3}$ and Joyjit Kund$\mbox{u}^{2}$}\\
{ \small
~\\
\noindent $^1$Laboratoire Charles Coulomb (L2C), University of Montpellier, CNRS, Montpellier, France\\
$^2$Department of Chemistry, Duke University, Durham, North Carolina 27708, USA\\ \vspace{-0.25cm}
$^3$Department of Physics, Duke University, Durham, North Carolina 27708, USA}

\end{center}\renewcommand{\theequation}{S\arabic{equation}}
\renewcommand{\thefigure}{S\arabic{figure}}
\renewcommand{\thesection}{S\arabic{section}}

\setcounter{equation}{0}
\setcounter{section}{0}
\setcounter{figure}{0}

\setlength{\parskip}{0.25cm}%
\setlength{\parindent}{0pt}%

\section{Simulation details and Model Parameters}
We simulate systems of hard spheres in $d=2,\dots,8$ with continuous size dispersity (polydispersity) using Monte Carlo (MC) simulations with a constant number of particles ($N=2000$ for $d\leq6$ and $N=7000$ in $d=8$), and a constant volume under periodic boundary conditions. For a given size distribution $P(\sigma)$, the mean diameter $\langle \sigma \rangle$ defines the unit of length and the standard deviation of that distribution, $\Delta=\sqrt{\langle \sigma^2 \rangle-\langle \sigma \rangle^2}$, defines the degree of polydispersity. 

We perform both the standard and swap MC (SWAP) dynamics. The standard MC protocol consists solely of translational displacement moves, uniformly drawn over a $d$ dimensional hypercube of side $\ell(d)$. The equilibrium fluid configurations deep inside the glassy regime are obtained using SWAP that involves both local displacements and non-local particle swaps, in which two randomly selected particles exchange their diameters. Particle swaps and displacements are attempted with probability $0.2$ and $0.8$, respectively, and are accepted if no overlap results. For a given $d$, the value of $\ell$ is chosen such that the relaxation time (in units of MC sweeps) for the standard dynamics is minimal near the dynamical transition (see \f{figS1}) (a). A large value of $\ell$ leads to unsuccessful displacement attempts, and a small value is inefficient at sampling the particle cage. We find $\ell=0.110$, $0.060$, $0.040$, $0.033$, $0.027$, and $0.021$ to be optimal in $d=2$, $3$, $4$, $5$, $6$, and $8$, respectively. These values are robust against the degree of polydispersity. Swap moves attempt to exchange the diameters of two particles with diameter difference $<d\sigma_{\rm tol}$, which roughly corresponds to the cage diameter. We also optimize the value of $d\sigma_{\rm tol}$ in different dimensions-- a representative plot for $d=6$ is shown in \f{figS1}) (b). We set $d\sigma_{\rm tol}=0.10$, $0.09$, $0.09$, $0.04$, $0.03$, and $0.012$ in $d=2$, $3$, $4$, $5$, $6$ and $8$ respectively. Please note that SWAP efficiency depends only weakly on $d\sigma_{\rm tol}$ in the range $0.01\leq d\sigma_{\rm tol}\leq 0.10$.

\section{Thermalization: Dynamic and static observables}
To ensure thermalization at each state point, system are evolved at least up to $20\tau_\alpha^{\rm swap}$ (measured from the decay of the overlap function-- see \f{figS2}), before starting the production run. In order to measure dynamical observables at relatively low packing fractions, production runs lasting at least $40 \tau_\alpha$ are used to average over time, while at high densities, $60$ replicas are run up to a shorter time, typically $\sim 5\tau_\alpha$, and the overall results are averaged. Typical correlation decays are given in \f{figS2}. We consider the dynamics starting from the onset of glassiness, which is detected by the emergence of  a inflection point in the mean squared displacement, thus implying non-Fickian diffusion (see \f{figS3}). The corresponding $\tau_0$ is estimated by the relaxation time at the onset for both the standard and the swap dynamics. We obtain $\tau_0^{\rm std}\approx 1400$ and $\tau_0^{\rm swap}\approx 2400$ in $d=2$; $\tau_0^{\rm std}\approx 1800$ and $\tau_0^{\rm swap}\approx 3000$ in $d=3$; $\tau_0^{\rm std}\approx 2700$ and  $\tau_0^{\rm swap}\approx 4200$ in $d=4$; $\tau_0^{\rm std}\approx 4000$ and $\tau_0^{\rm swap}\approx 5600$  in $d=5$; $\tau_0^{\rm std}\approx 5000$ and $\tau_0^{\rm swap}\approx 6500$ in $d=6$; $\tau_0^{\rm swap}\approx 6500$ and $\tau_0^{\rm swap}\approx 8200$ in $d=8$.

Various static observables, such as the structure factor and the pair correlation function, are used to detect putative (and unwanted) crystallization. An instance of fractionation is shown in \f{figS4}. The pressure, $P$, is extracted from the contact value of the pair correlation function properly scaled for a polydisperse system, to calculate the equation of state, $Z(\phi)=P/\rho$, where $\phi=\bar{V}_d\rho$ is the packing fraction for number density $\rho$ and average sphere volume $\bar{V}_d$\c{yuste2005}. Equations of states for different $\Delta$ and $d$ are shown in \f{figS5}. 
In \f{figS6}, we show SWAP efficiency as a function of polydispersity for different values of sluggishness, given by $\tau_\alpha^{\rm std}$. To estimate the SWAP efficiency beyond the numerically accessible regimes, we use parabolic fitting forms extrapolating the data-- see \f{figS7}.
\begin{figure}
\includegraphics[width=\linewidth]{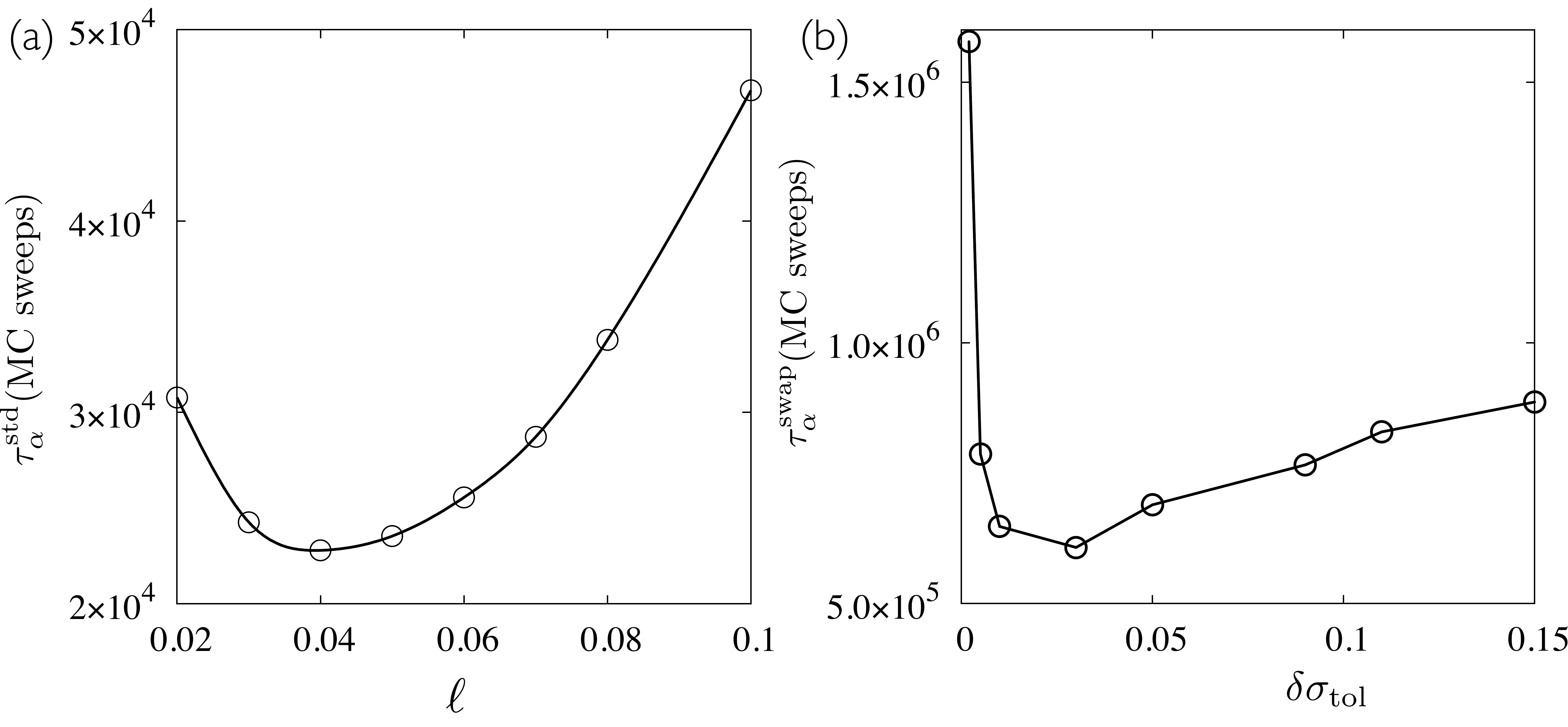}
\caption{(a) The structural relaxation time, $\tau_\alpha^{\rm std}$, for the standard dynamics is minimimal at a finite $\ell$, as shown here for $d=4$, $\Delta=10\%$, and $\varphi=0.4015$. (b) The relaxation time $\tau_\alpha^{\rm swap}$ for SWAP dynamics is minimal for a finite $\delta\sigma_{\rm tol}$, as shown here for $d=6$, $\Delta=10\%$, and $\varphi=0.1865$.}
\label{figS1}
\end{figure}
\begin{figure}
\includegraphics[width=\linewidth]{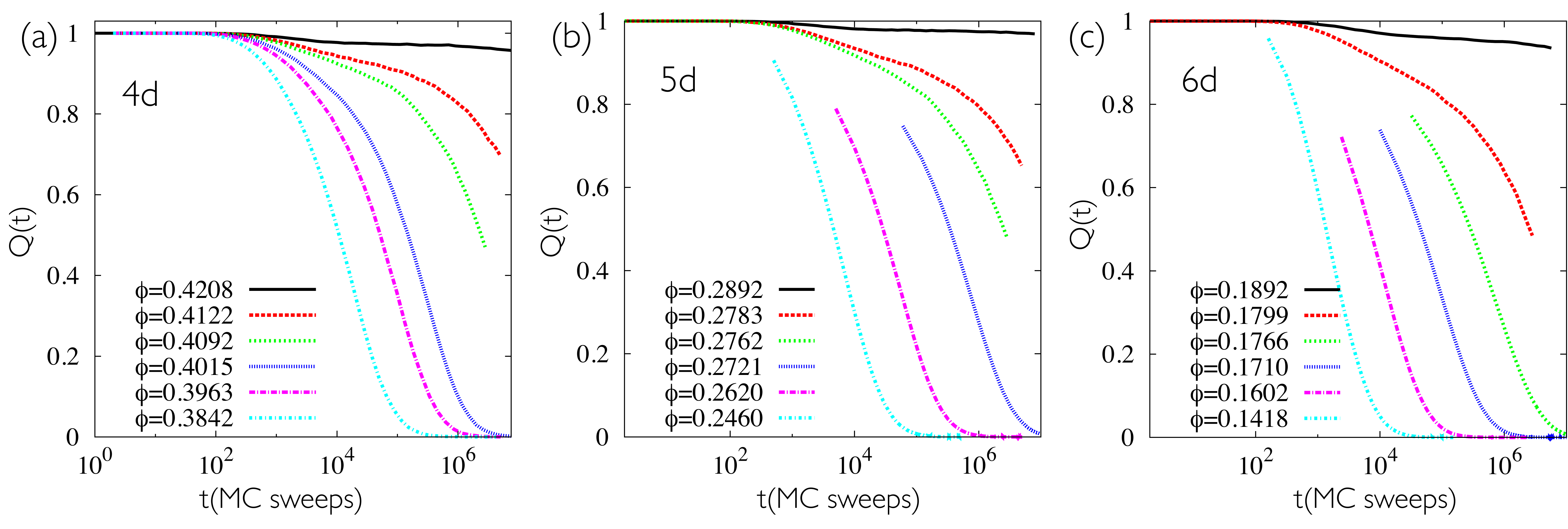}
\caption{Decay of the self-part of the overlap function $Q(t)$ (Eq.~1 in the main text) as a function of time for various densities in $d=4$, $5$, and $6$ for $\Delta=10\%$.  The $1/e$ decay of this function implicitly defines $\tau_\alpha$.}
\label{figS2}
\end{figure}
\begin{figure}
\includegraphics[width=\linewidth]{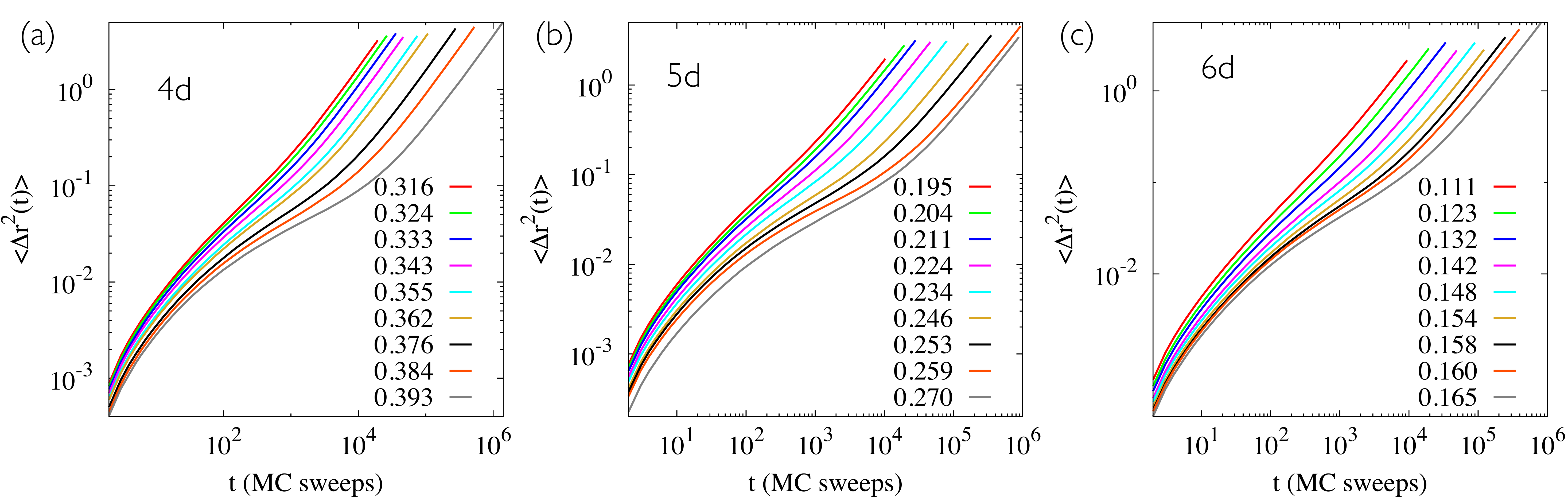}
\caption{Mean-squared displacement at different densities around the onset of glassiness, at which non-Fickian diffusion sets in. Results are shown here for the standard dynamics in (a) $d=4$ ($\phi_0\approx0.355$), (b) $d=5$ ($\phi_0\approx0.236$), and (c) $d=6$ ($\phi_0\approx0.152$). }
\label{figS3}
\end{figure}
\begin{figure}
\includegraphics[width=0.84\linewidth]{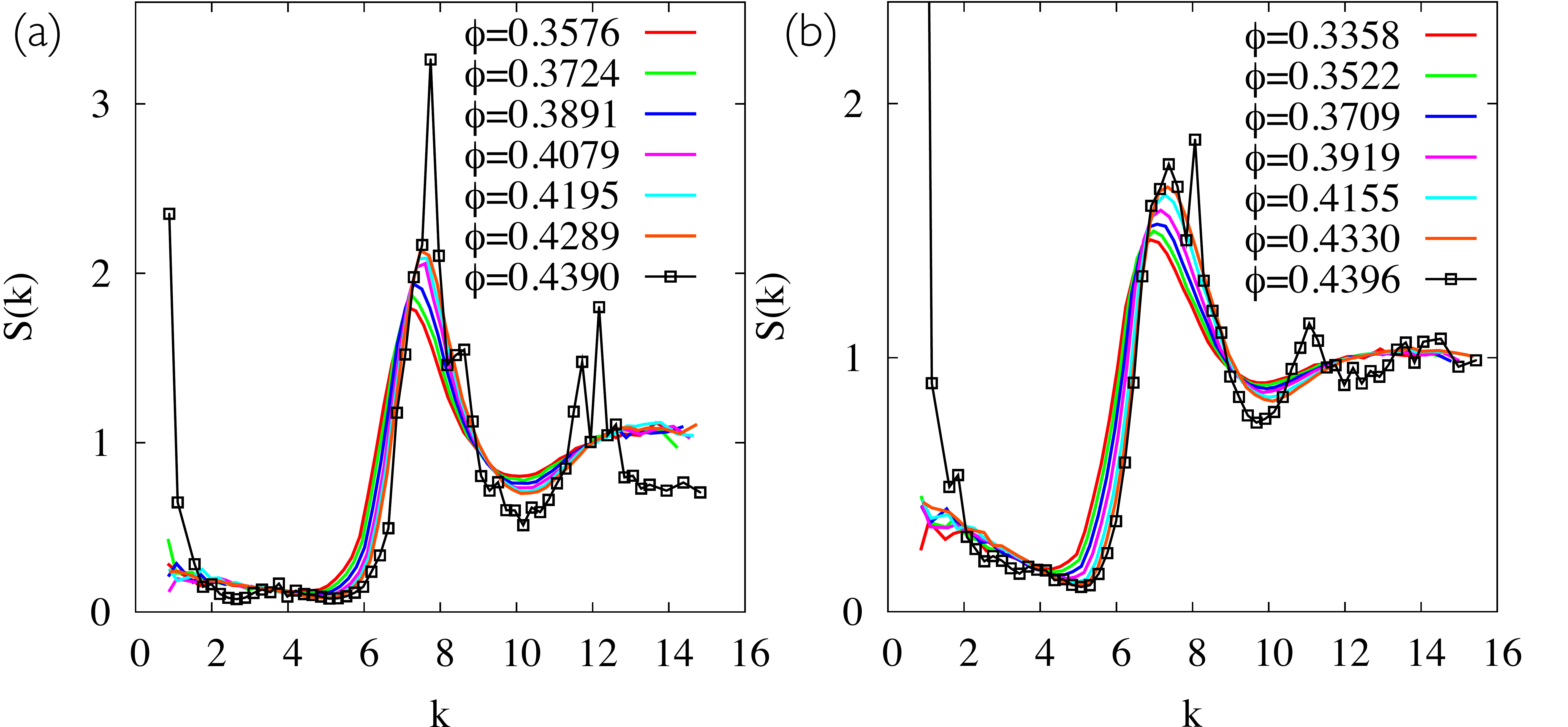}
\caption{Typical plots of the structure factor, $S({\rm k})$ for (a) $\Delta=15\%$, and (b) $\Delta=20\%$ in $d=4$ at various densities. For $\phi \gtrsim 0.43$, fractionation takes place, as indicated by a low k peak.}
\label{figS4}
\end{figure}
\begin{figure}[H]
\includegraphics[width=\linewidth]{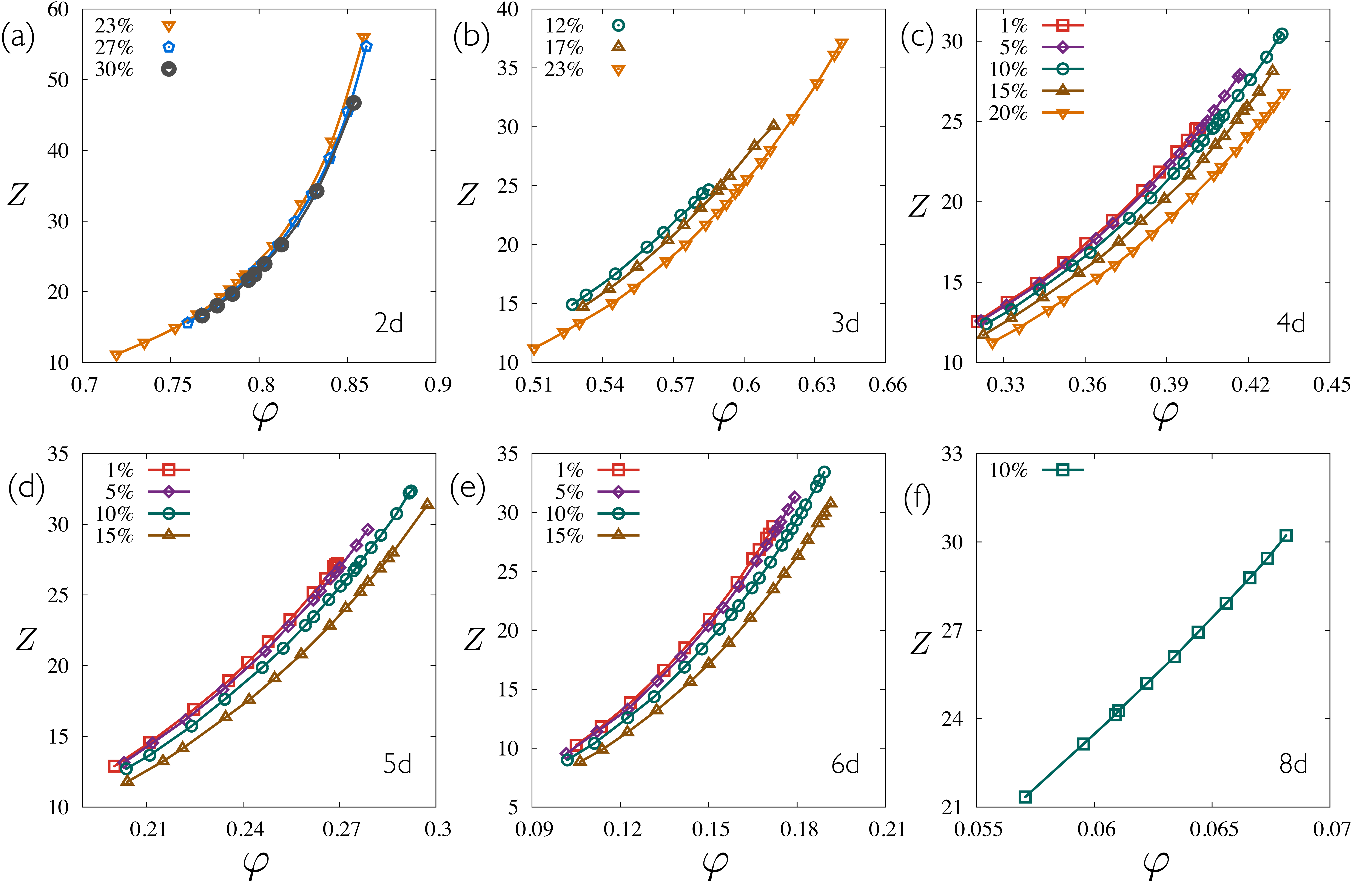}
\caption{Equations of states in $d=2$, 3, 4, 5, 6, and 8.}
\label{figS5}
\end{figure}
\begin{figure}[H]
\includegraphics[width=\linewidth]{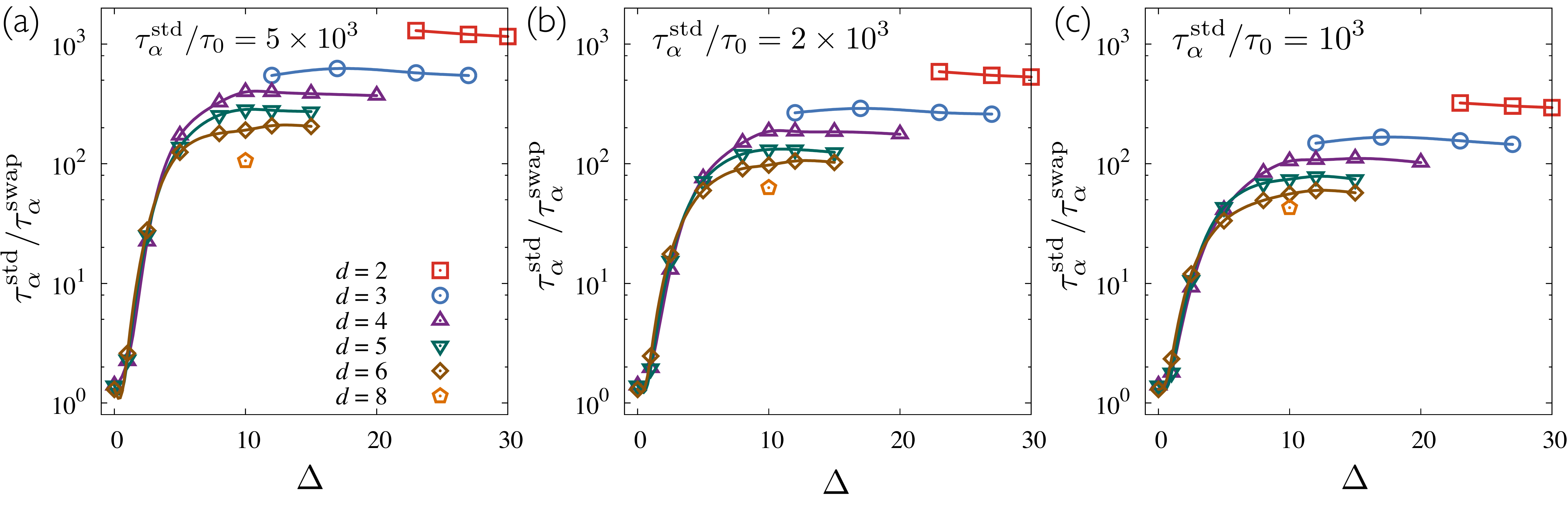}
\caption{The plateau height, which is the maximum efficiency for a given $d$, increases monotonically with $\tau_\alpha^{\rm std}/\tau_0$. Here we specifically consider $\tau_\alpha^{\rm std}/\tau_0=5\times 10^3$, $2\times 10^3$, and $10^3$, but the dimensional trend is robust against this choice.}
\label{figS6}
\end{figure}
\begin{figure}
\includegraphics[width=\linewidth]{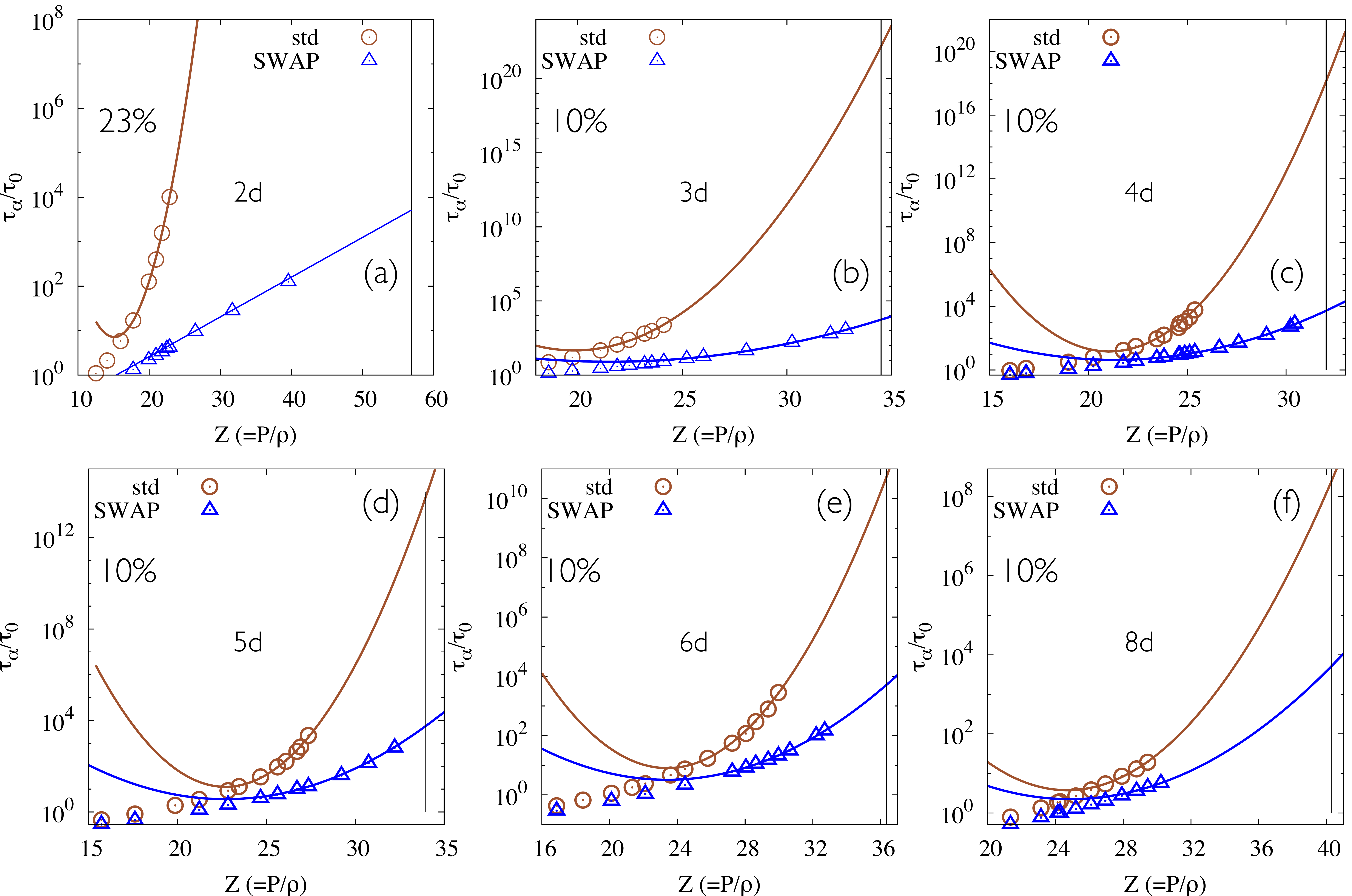}
\caption{The relaxation time $\tau_\alpha$ versus the reduced pressure $Z$ in different dimensions. The data for $\tau_\alpha$ are extrapolated using the parabolic form: $\tau=\tau_{\infty}\exp[A(Z-Z_0)^2]$. The vertical line corresponds to $\tau_\alpha^{\rm swap}/\tau_0=5\times 10^3$.}
\label{figS7}
\end{figure}
 \begin{figure}
\includegraphics[width=1.01\linewidth]{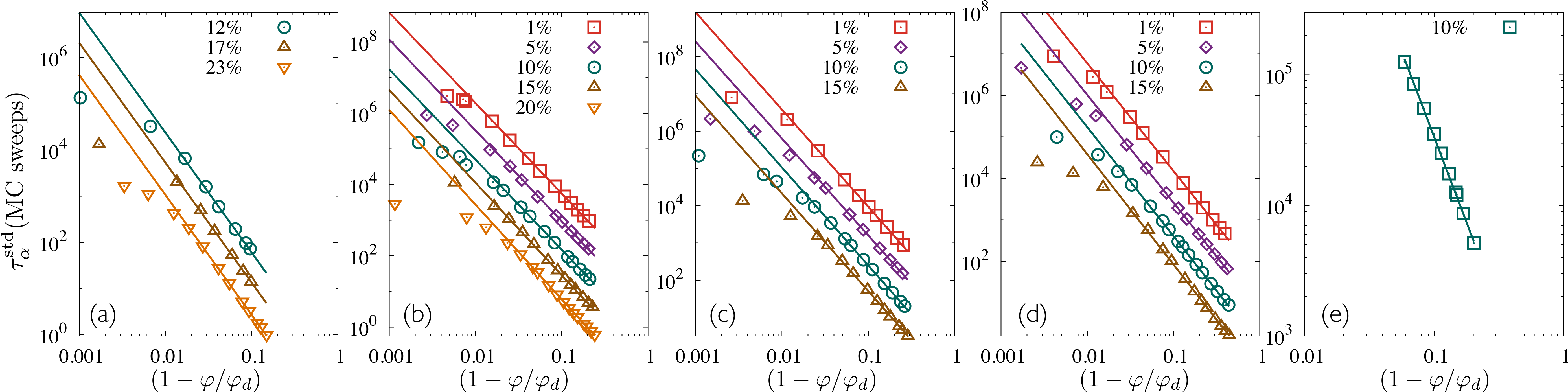}
\caption{Critical scaling of the relaxation time $\tau_\alpha^{\rm std}$ near the dynamical transition in (a) $d=3$, (b) $d=4$, (c) $d=5$, (d) $d=6$, and (e) $d=8$.}
\label{figS8}
\end{figure}
\begin{figure}
\includegraphics[width=1.01\linewidth]{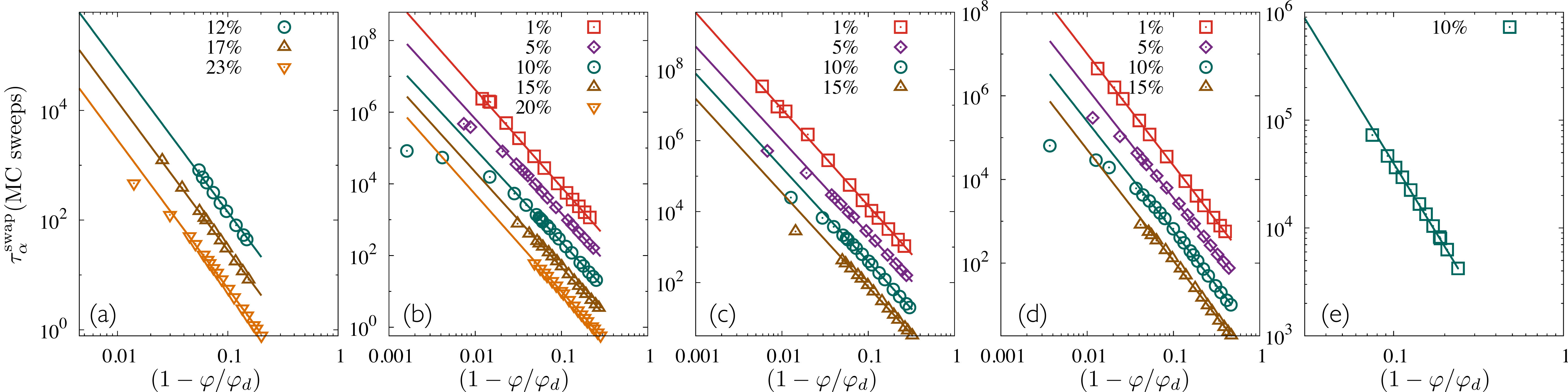}
\caption{Critical scaling of the relaxation time $\tau_\alpha^{\rm swap}$ near the dynamical transition in (a) $d=3$, (b) $d=4$, (c) $d=5$, (d) $d=6$, and (e) $d=8$}
\label{figS9}
\end{figure}
\section{Mode-coupling analysis}
In the limit $d\rightarrow\infty$, there exists a dynamical critical point, $\varphi_\mathrm{d}$, at which the relaxation time $\tau_\alpha$ diverges as $\sim (\phi_d-\phi)^{-\gamma}$ and the system gets trapped in one of the many metastable minima. This dynamical criticality is avoided in finite dimensions because of competition from activated processes. One can nonetheless fit a critical form to $\tau_\alpha$ versus $\phi$ over a limited regime below $\varphi_\mathrm{d}$ to estimate $\varphi_\mathrm{d}$ and $\gamma$ (see \f{figS8} and~\ref{figS9}). This power-law fit becomes more and more accurate with increasing $d$. Our estimates of $\varphi_\mathrm{d}$, and $\gamma$ for both the standard and swap dynamics in different dimensions and for different values $\Delta$ are given in Table~I and II. 

\begin{table} 
  \begin{tabular}{|p{1.1cm}|p{1.1cm}|p{1.1cm}|p{1.1cm}|p{1.1cm}|p{1.1cm}|p{1.1cm}|p{1.1cm}| p{1.1cm}|p{1.1cm}|p{1.1cm}|p{1.1cm}|p{1.1cm}|p{1.1cm}|p{1.1cm}|}
 \hline
  \multicolumn{3}{|c|}{$d=3$} &
 \multicolumn{3}{|c|}{$d=4$} &
 \multicolumn{3}{|c|}{$d=5$} &
  \multicolumn{3}{|c|}{$d=6$} &
  \multicolumn{3}{|c|}{$d=8$} \\
 \hline
 $\Delta$ & $\phi_\mathrm{d}^{\rm std}$  & $\gamma$ & $\Delta$ & $\phi_\mathrm{d}^{\rm std}$& $\gamma$   & $\Delta$ & $\phi_\mathrm{d}^{\rm std}$& $\gamma$  & $\Delta$ & $\phi_\mathrm{d}^{\rm std}$& $\gamma$  & $\Delta$ & $\phi_\mathrm{d}^{\rm std}$& $\gamma$\\
 \hline
 12\%& 0.5830 & 2.60 & 1\% & 0.4038 & 2.54 & 1\% & 0.2688 & 2.61 & 1\% & 0.1725 & 2.60 & - & - & - \\
 17\%& 0.5895 &  2.60 & 5\% & 0.4051 & 2.55 & 5\% & 0.2705 & 2.60 & 5\% & 0.1747 & 2.60 & - & - & -\\
 23\%& 0.6000 & 2.60 &10\% & 0.4101 & 2.53 &10\% & 0.2769 & 2.63 & 10\% & 0.1807 & 2.63 & 10\% & 0.07157 & 2.61\\
-& - & -& 15\% & 0.4181 & 2.60 & 15\% & 0.2862 & 2.62 & 15\% & 0.1899 & 2.66 & - & - &-\\
 \hline
\end{tabular}
 \label{table01}
 \caption{The dynamical transition density $\phi_\mathrm{d}^{\rm std}$ and the corresponding critical exponent $\gamma$ for different $\Delta$ in $d=3$, $4$, $5$, $6$ and $8$ for the standard dynamics.}
 \vspace{0.5 cm}
  \begin{tabular}{|p{1.1cm}|p{1.1cm}|p{1.1cm}|p{1.1cm}|p{1.1cm}|p{1.1cm}|p{1.1cm}|p{1.1cm}| p{1.1cm}|p{1.1cm}|p{1.1cm}|p{1.1cm}|p{1.1cm}|p{1.1cm}|p{1.1cm}|}
 \hline
  \multicolumn{3}{|c|}{$d=3$} &
 \multicolumn{3}{|c|}{$d=4$} &
 \multicolumn{3}{|c|}{$d=5$} &
  \multicolumn{3}{|c|}{$d=6$} &
  \multicolumn{3}{|c|}{$d=8$} \\
 \hline
 $\Delta$ & $\phi_\mathrm{d}^{\rm swap}$  & $\gamma$ & $\Delta$ & $\phi_\mathrm{d}^{\rm swap}$& $\gamma$   & $\Delta$ & $\phi_\mathrm{d}^{\rm swap}$& $\gamma$  & $\Delta$ & $\phi_\mathrm{d}^{\rm swap}$& $\gamma$  & $\Delta$ & $\phi_\mathrm{d}^{\rm swap}$& $\gamma$\\
 \hline
 12\%& 0.6189 & 2.70 & 1\% & 0.4068 & 2.72 & 1\% & 0.2711 & 2.72 & 1\% & 0.1741 & 2.67 & - & - & - \\
 17\%& 0.6286 &  2.71 & 5\% & 0.4200 & 2.64 & 5\% & 0.2807 & 2.64 & 5\% & 0.1813 & 2.62 & - & - & -\\
 23\%& 0.6399 & 2.71 &10\% & 0.4331 & 2.59 &10\% & 0.2915 & 2.65 & 10\% & 0.1899 & 2.62 & 10\% & 0.07506 & 2.57\\
-& - & -& 15\% & 0.4426 & 2.66 & 15\% & 0.3017 & 2.65 & 15\% & 0.1997 & 2.66 & - & - &-\\
 \hline
\end{tabular}
\caption{The dynamical transition density $\phi_\mathrm{d}^{\rm swap}$ and the corresponding critical exponent $\gamma$ for different $\Delta$ in $d=3$, $4$, $5$, $6$ and $8$ for the SWAP.}
\label{table01}
\end{table}

\end{document}